\documentclass[review,12pt]{elsarticle1}
\usepackage{lineno}
\usepackage[colorlinks=true, urlcolor=blue, pdffitwindow=true, linkcolor=blue, citecolor=blue, pdfstartview={FitH}]{hyperref}
\usepackage{bm, amsmath, amssymb}
\usepackage{epstopdf, graphics}
\usepackage{multirow}
\usepackage{longtable}
\usepackage[utf8]{inputenc}
\usepackage{url}
\usepackage{numcompress}
\usepackage[bitstream-charter]{mathdesign}
\biboptions{sort&compress}
\usepackage[twoside,paperwidth=210mm, paperheight=297mm, textheight=682pt,
textwidth=480pt,
centering,
headheight=50pt,
headsep=12pt,
footskip=18pt,
footnotesep=24pt plus 2pt minus 12pt,
columnsep=18pt]{geometry}
\journal{}
\makeatletter
\def\ps@pprintTitle{%
  \let\@oddhead\@empty
  \let\@evenhead\@empty
  \def\@oddfoot{\reset@font\hfil\thepage\hfil}
  \let\@evenfoot\@oddfoot
}
\makeatother
\begin{document}

\begin{frontmatter}
\title{ Uniform magnetic field on the relativistic spinless particles with constant rest mass: 2D polar space}

\author[1,2]{Sami Ortakaya}
\ead{sami.ortakaya@yahoo.com}
\address[1]{Ercis Central Post Office, 65400 Van, Turkey}
\address[2]{Shipito address: 444 Alaska Avenue Suite $\#$BKF475
Torrance, CA 90503 USA}
\begin{abstract}
{\footnotesize We present an interaction modeling for the relativistic spin-0 charged particles moving in a uniform magnetic field. In the absence of the improved perturbative way, we solve directly Kummer\rq{}s differential equation including principal quantum numbers. As a functional approach to the nuclear interaction, we consider particle bound states without antiparticle regime. 
Within the approximation line to $1/r^4$, we have also improved the considerations of the $V(r)$$\neq$$0$ and $S(r)$$=$$0$ related to scalar and mass interactions. Moreover, we have founded a closeness for introduced approximation scheme for range of $0.5$ and $1.0$ $\mathrm {fm}$. In this way, minimal coupling might also yields analytically energy spectra. Within the spin-zero relativistic regime, we have considered the inverse-square interaction under uniform magnetic field and we have also founded that the energy levels increase with increasing interaction energy (i.e, quantum well width decreases for given values). Additionally, energy levels increase with larger values of the uniform magnetic fields. The charge distributions is also valid for the central interaction-confinement space. Putting the approximation to spin-zero motion with $V(r)$$\neq$$0$ and $S(r)$$=$$0$, one can introduced solvable model in the 2D polar space.}
\end{abstract}


\end{frontmatter}
%{\color{blue}\today}
%\linenumbers
\section{Introduction}\label{sec1}

Spinless quantum states of the relativistic Klein-Gordon equation might view an effective modeling in the particle physics. Considering interaction field, it can be taken relativistic output related to the quantum states. The empirical and solvable programs under the polynomial regime for the subatomic placement might also become a toy model. In a way, both scalar interaction and spin symmetries has been accomplished within the spin-zero and spinor forms, respectively, in which mathematical usage is valid \cite{Jia2009, Castro2019, Berkdemir2006, sami}. Especially, considering that rest mass with spatial formation $m_{0}c^2+S(r)$ cause to solvable models for the relativistic eigenvalues, analytical aspects yield acceptable quantum wavefunctions for the equality which describe charge interaction and mass distribution, so many authors have studied on this \cite{biswas, saad, QIANG20084789}. As can be seen in the several works, the basis of the spatial dependent mass is that solvable model of the Klein-Gordon equation is realized by taking $S(r)$=$V(r)$. Not only empirical potential energies but also spatial varying magnetic field acts on the spin-0 quantum system; nevertheless, the equality between radially varying interaction $V(r)$ and spatial dependent mass-energy $S(r)$ has been considered Killingbeck formation under Aharonov-Bohm flux \cite{ikhdair}. Within 1D space, exponential form of the mass distribution through zero interaction $V(r)$=0, has been also tried under magnetic field with the same form \cite{aydogdu}. Furthermore, external magnetic field appears on the spin-0 states including the familiar equality for shape invariant method \cite{ctp}. As another exact solution procedure, relativistic motion of the spin-0 particle has been analyzed under uniform electric and magnetic fields as well \cite{liu}.

As a mathematical modeling of the spin-0 relativistic quantum confinement, in the presence of the \lq\lq{}pure functional usage\rq\rq{}, iterative procedure exists on the polynomial representation \cite{Ciftci2003}. Likewise, symmetry transform \cite{Fernandez2004} and algorithm steps \cite{Mielnik2000} construct acceptable solutions. Regarding new space transform of the radial distribution, quantum number dependencies have also been accomplished considering terminal value through the Laplace transform \cite{Ortakaya2013a}. In recent study, the analysis into relativistic anharmonicity has used to be analytical discussion of the Kummer\rq{}s orthogonality \cite{Ortakaya2013b}.

Because of the mathematical backgrounds permit orthogonal eigenvalue systems, relativistic pion-like polynomial network can be considered via a uniform magnetic field. Rather than the spatial distribution of mass-energy which equals to central interaction, spin-zero particle states with constant rest mass exist on the improved approximation. As will seen, we only get a functional procedure near fixed distance in the considered approximation. 

The quantum system without antiparticle context may denotes only spinless \lq\lq{}particle\rq\rq{} levels. In this way, we deal with energy spectra of the constant rest mass. Due to the uniform magnetic field on the 2D polar system, it can also be obtained new changes in the energy states adding magnetic field terms. Besides rest mass without spatial distribution cause to square of the spatial distribution, by expanding series of the $V(r)^2$ at fm-scale, the deep analytical treatment gives reduced variables within solvable space, so spin-0 particle states with constant mass has been considered in the 2D polar coordinate space. Although the equality between \lq\lq{}scalar and vector potentials\rq\rq{} gives acceptable ways near a given distance, we can apply nuclear distance to the particle states without antiparticle regime through $S(r)$=0 and also consider charged relativistic particles which interact with stopping material via inverse square potential by $V(r)$$\neq$0. From the energy spectra we understand that the decreasing in the energy levels with increasing of the quantum well width denotes only particle regime without antiparticle sea. 

\section{Theoretical Model}

We present a quantum system including relativistic spin-0 charged particles. In order to get an energy spectra, we consider relativistic charges expose to the attractive forces modeled by subatomic interaction, so we focus on he relativistic particle through central inverse square potential energy. In the atomic units ($\hbar=c=e=1$), we get the Klein-Gordon equation for a charged particle of rest mass $m_{0}$, energy-eigenvalues can be taken from \cite{Greiner}
\begin{equation}
\left(\mathrm{i}\frac{\partial}{\partial t}-A_{0}\right)^2 \Psi(\vec{\bm{r}})=\left(\left(\bm{p}-{\rm\textbf  {A}}\right)^2+m_{0}^2\right)\Psi(\vec{\bm{r}}),
\end{equation}
where $\bm{p}$ denotes the momentum defined as covariant derivative $-\mathrm{i}\bm{\nabla}$ and $A_{\mu}$ is the four-vector potential. We also consider that first component $A_{0}=V(r)$ defines interaction and apply a uniform magnetic field $\rm\textbf  {A}$$\neq$0. Rest mass energy might has a radial form given as $m_{0}c^2\to m_{0}c^2+S(r)$. Assuming $S(r)$=0 with Coulomb gauge $ \bm{\nabla}\cdot \rm\textbf  {A}$=0, Klein-Gordon equation reads
\begin{equation}\label{e2}
\left(-\bm{\nabla}^2 + {\rm i}2{\rm \textbf{A}}\cdot\bm{\nabla} + m_{0}^2 \right) \Psi(\vec{\bm{r}})=\left(E-V(r)\right)^2\Psi(\vec{\bm{r}}).
\end{equation}
The vector potential within considered gauge is derived from applied magnetic field $\rm \textbf{B}$=$B_{0} \rm \hat{\textbf{z}}$   (perpendicular to confinement plane), so vector potential has the form
$${\rm \textbf{A}}={\rm \textbf{B}}\times {\rm \textbf{r}},\quad {\rm \textbf{A}}=B_{0}r\bm{\hat{\phi}}$$ In practice, we refer to 2D polar eigenfunctions derived from $$\Psi(\vec{\bm{r}}:\,r,\,\phi)= \frac{1}{\sqrt{2\pi}}u_{n m}(r){\rm{e}}^{{\rm{i}} m\phi},$$ 
where $m=0,\,\pm1,\,\pm2,\,\pm3\dots\;\; {\rm and}\;\; n=0,\,1,\,2,\,3\dots$

\section{Approximate Solutions to $\frac{1}{r^4}$}
\begin{figure*}[!hbt]\center
\scalebox{1.2}{\includegraphics{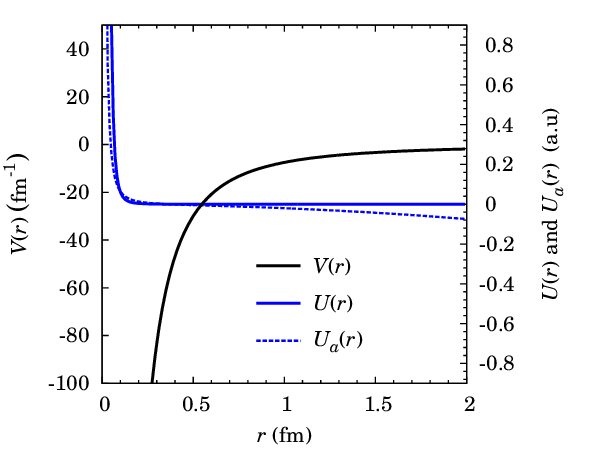}}
\caption{Attractive inverse-square potential and its approximation near $r_0$=0.5 fm through $Z_{0}$=30 $\rm fm^{-1}$.}\label{f1}
\end{figure*}

The central potential energy for spin-0 particles which interact with a stopping material can be defined by inverse-square function of the form
\begin{equation}\label{pot}
V(r)=-\frac{Z_{0}r_{0}^2}{r^2},
\end{equation}
where $Z_{0}$ is the \lq\lq{}quantum well\rq\rq{} width (in $\rm fm^{-1}$ unit) and $r_{0}$ denotes the nuclear placement for a given fixed distance. Putting Eq. (\ref{pot}) into Eq. (\ref{e2}), we rewrite radial part
\begin{eqnarray}\label{e44}
u''(r)+\frac{1}{r}u'(r)+\big[\left(E_{nm}-V(r)\right)^2- m_{0}^2-\frac{m^2}{r^2}+2mB_{0}-B_{0}^2 r^2\big]u(r)=0,\nonumber\\
{} \end{eqnarray}
using a new variable $x=\frac{r^2}{r_0^2}$, we should have
\begin{equation}\label{maine}x u\rq{}\rq{}(x)+u\rq{}(x)+\left[\beta_{1}^2-\beta_{2}^2 x-\frac{\kappa^2 }{x} +U(x)\right]u(x)=0,
\end{equation}
where
\begin{eqnarray}
\begin{aligned}
\beta_{1}^2&=\frac{r_{0}^2}{4}\bigg(E_{nm}^2-m_{0}^2+2mB_{0}\bigg),&&\\
\beta_{2}^2&=\frac{B_{0}^2r_{0}^4}{4},&&\\
\kappa^2&=\frac{r_{0}^2}{4}\left(\frac{m^2}{r_{0}^2}-2Z_{0}E_{nm}\right)
\end{aligned} 
\end{eqnarray}
and
\begin{equation}
U(x)=\frac{ Z_{0}^2 r_ {0}^2}{4 x^2}.
\end{equation} 
In order to solve Eq. (\ref{maine}) analytically, we shall use a new series approximation to $\frac{1}{x^2}$. Then, we consider that the treatment becomes in the form
\begin{equation}
U(x)\simeq U_{a}(x)=\frac{r_{0}^2}{4}\left[A_{1}+A_{2}x+\frac{A_{3}}{x}\right]
\end{equation}
as an improved approximation. Expanding $U(r)$ and $U_{a}(r)$ as series of $r$ near $r=r_{0}\,(x=1)$ and comparing up to $3^{\rm rd}$ order, we take that
\begin{equation}
A_1 =-3Z_{0}^2,\qquad A_2 =Z_{0}^2,\qquad A_3=3Z_{0}^2 .
\end{equation}
One can see from Fig. \ref{f1} that the spatial behaviors related to $U(r)$ and $U_{a}(r)$ show a good closeness at the nuclear distance $r_{0}$=0.5 fm. Also, attractive interaction of the values $Z_{0}$=30 to 50 $\rm fm^{-1}$ provide this agreement.

Putting the new approximation into Eq. (\ref{maine}), the solvable equation reads
\begin{equation}\label{maine2}
x \overline{u}\rq{}\rq{}+\overline{u}\rq{}+
\bigg[-\overline{\beta}_{1}^2-\overline{\beta}_{2}^2 x-\frac{\overline{\kappa}^2} {x}\bigg]\overline{u}=0,
\end{equation}
where
\begin{eqnarray}
\begin{aligned}
\overline{\beta}_{1}^2&=\frac{r_{0}^2}{4}\bigg(E_{nm}^2-m_{0}^2+2mB_{0}+A_{1}\bigg),&&\\
\overline{\beta}_{2}^2&=\frac{r_{0}^2}{4}\left(B_{0}^2r_{0}^4-A_{2}\right),&&\\
\overline{\kappa}^2&=\frac{r_{0}^2}{4}\left(\frac{m^2}{r_{0}^2}-2Z_{0}E_{nm}-A_{3}\right).
\end{aligned} 
\end{eqnarray}
Letting $\overline{u}(x)$ provided to be $\overline{u}(x)=x^{-\overline{\kappa}}f(x);$ for $\overline{\kappa}>0$, Eq. (\ref{maine2}) yields
\begin{equation}\label{maine3}
x f\rq{}\rq{}(x)+(1-2\overline{\kappa})f\rq{}(x)+
\big[\overline{\beta}_{1}^2-\overline{\beta}_{2}^2 x\big]f(x)=0.
\end{equation}
Note that the wavefunction leads to as $x\to0$ $(r\to 0)$, so physical regime has to be $f(x)\propto x^{\sigma};$ via $\sigma>\overline{\kappa}$.
Then, one can see from Laplace transform that, Eq. (\ref{maine3}) yields $n$\rq{}th eigenvalue spectra \cite{Ortakaya2013a}
\begin{equation}\label{nn}
\frac{\overline{\beta}_{1}^2}{2\overline{\beta}_{2}}-\overline{\kappa}=n+\frac{1}{2},\qquad(n=0,\,1,\,2\,\dots)
\end{equation}
and Kummer\rq{}s solution leads to unnormalized form
\begin{equation}
\overline{u}(x)=\rho^{\overline{\kappa}}{\rm e}^{-\overline{\beta}_{2}x}M\left(-n,\,2\overline{\kappa}+1;\,2\overline{\beta}_{2}x\right).
\end{equation} 
\section{Numerical Results}
Solving the energy-spectra equation given in Eq. (\ref{nn}), it can be concluded that the energy levels increase with increasing principal quantum number $n$. One can see from transcendental equation at real values, the upper levels approach zero-level with increasing quantum numbers. In that case, solutions of the particle states have bound state levels while anti-particle states do not exist in the atomic units. 
\begin{figure}[!hbt]
\centering
\scalebox{1.2}{\includegraphics{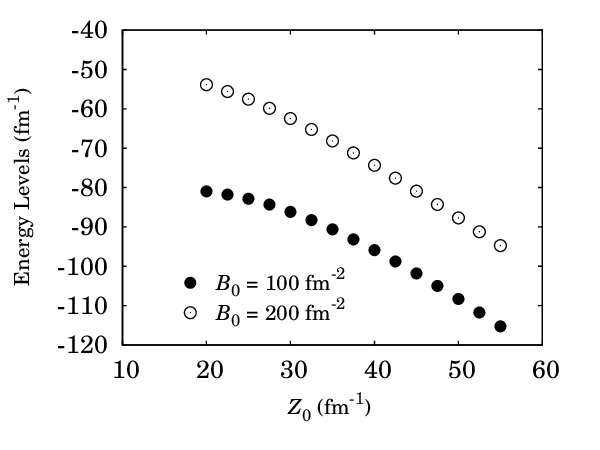}}\\
\scalebox{1.2}{\includegraphics{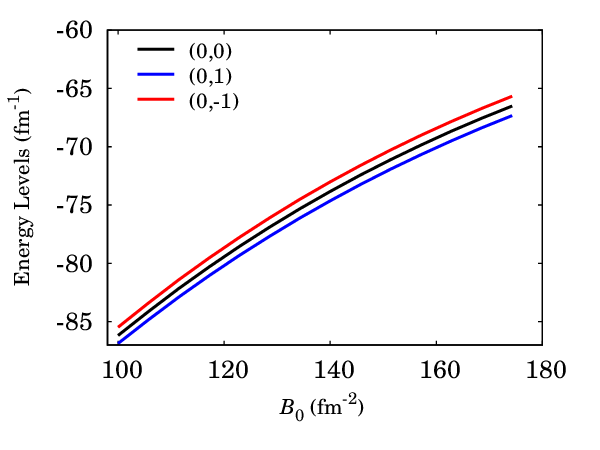}}
\caption{Relativistic energy levels for inverse-square interaction when rest mass energy and nuclear distance taken as 120 $\rm fm^{-1}$ and 1.0 fm, respectively. }\label{d0}
\end{figure}
Fig. \ref{d0} shows eigenvalues ($n$=0, $m$=0) as a function of potential height parameter $Z_{0}$ at atomic scale of $m_{0}$=120 $\rm fm^{-1}$. In the presence of the magnetic field, the results are valid for fixed values of magnetic field 100 and 200 $\rm fm^{-2}$. Moreover, it can be easily obtained that particle levels approach to zero-level with smaller well width (i.e.,  decreasing $Z_{0}$) which refers to narrow well. Note that potential-height has also the validity on the considered approximation at the range of $20-50\,\rm fm^{-1}$. Applying the functional procedure in Eq. (\ref{nn}), relativistic eigenvalues decrease at the ranges $-120\, {\rm fm^{-1}}<E_{nm}<-90\, {\rm fm^{-1}}$ through particle levels.  

\begin{figure}[!hbt]\center
\scalebox{1.2}{\includegraphics{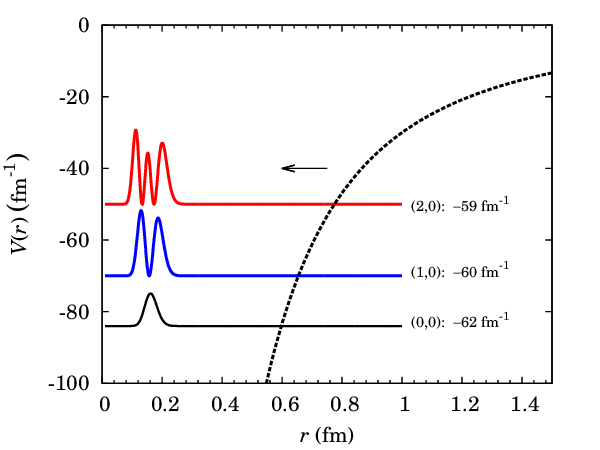}}
\caption{Charge densities for the relativistic spin-0 particles moving in the magnetic field, $B_{0}$=$200\, \rm fm^{-2}$ denoted by arbitrary ($n,\,m$) states.}\label{wp}
\end{figure}

Nevertheless, the relativistic energies as a function of $B_{0}$ behave a linear interpolation at smooth energy-ranges through $-90\, {\rm fm^{-1}}<E_{nm}<-60\, {\rm fm^{-1}}$. Besides bound-state levels increase with increasing magnetic field, larger energy values exist for $m$=$-1$ while smaller values obtained for $m$=1. Rather than antiparticle regime, decreasing in the binding energies shows that particle states are expected as $E$$<$0. Then, bound states for spin-0 regime can be easily provided at given atomic scale. 

Now we get wave-functional behaviors for charge density given as \begin{equation}\varrho(r)=\frac{\pm E_{nm}}{m_{0}}u^{*}(r)u(r)\nonumber\end{equation}
Here the solutions which being for particle bound sates, given in Fig. \ref{wp}. When putting $Z_{0}$=$30\, \rm fm^{-1}$ at $r_0$$=$1.0 fm and varying $(nm)$  as $(00)$, $(10)$ and $(20)$, the charge distributions are localized near 0.2 fm.

\section{Conclusion}
Within the framework of the relativistic spin-zero charge through potential energy of the form $-1/r^2$, the quantum-eigenvalues have proved to be a scale at value of the well width 0.5 fm. Relating the results to the equality as $S(r)$=0, $V(r)$$\neq$0), bound states occur easily considering uniform magnetic field with increasing energy levels. The key role regarding relativistic spin-zero states is to find the improvement scheme to the planar confinement with $\rm 4^{th}$ order, so bound states of the relativistic charge without spatial-mass distribution are founded and acceptable eigenfunctions also showed as a \lq\lq{}charge density\rq\rq{}.
\section{Data Availability Statement}
No Data associated in the manuscript
\section{Conflict of Interest}
The author declares that he has no confict of interest.
\section{Acknowledgement}\label{secA1}
The author wishes to thank the M.Sc. Sefa Ortakaya for helpful library support.

\end{document}